\documentclass[twocolumn,showpacs,amsfonts,nofootinbib,aps,prc,floatfix,superscriptaddress]{revtex4-1}

\usepackage{mathrsfs}
\usepackage{epsfig}
\usepackage{amssymb}
\usepackage{amsfonts}
\usepackage{latexsym}
\usepackage{amsthm}

\usepackage{bm}
\usepackage{graphicx}
\usepackage{amsmath}
\usepackage{hyperref}
\usepackage{slashed}
\usepackage{soul}

\usepackage{ulem}
\usepackage{color}

\begin{document}


\title{Anisotropic flow of thermal photons as a quark-gluon plasma viscometer}

\author{Chun Shen}
\email[Correspondence:\ ]{chunshen@physics.mcgill.ca}
\affiliation{Department of Physics, McGill University, 3600 University Street, Montreal, Quebec, H3A 2T8, Canada}
\affiliation{Department of Physics, The Ohio State University,
  Columbus, Ohio 43210-1117, USA}
\author{Ulrich Heinz}
\affiliation{Department of Physics, The Ohio State University,
  Columbus, Ohio 43210-1117, USA}
\author{Jean-Fran\c{c}ois Paquet}
\affiliation{Department of Physics, McGill University, 3600 University Street, Montreal, Quebec, H3A 2T8, Canada}
\author{Igor Kozlov}
\affiliation{Department of Physics, McGill University, 3600 University Street, Montreal, Quebec, H3A 2T8, Canada}
\author{Charles Gale}
\affiliation{Department of Physics, McGill University, 3600 University Street, Montreal, Quebec, H3A 2T8, Canada}
\affiliation{Frankfurt Institute for Advanced Studies, Ruth-Moufang-Str. 1, D-60438 Frankfurt am Main, Germany}

\begin{abstract}
We present state-of-the-art calculations of viscous photon emission from nuclear collisions at RHIC and LHC. Fluctuating initial density profiles are evolved with event-by-event viscous hydrodynamics. Momentum spectra of thermal photons radiated by these explosively expanding fireballs and their $p_T$-differential anisotropic flow coefficients $v_n(p_T)$ are computed, both with and without accounting for viscous corrections to the standard thermal emission rates. Viscous corrections to the rates are found to have a larger effect on the $v_n$ coefficients than the viscous suppression of hydrodynamic flow anisotropies. The benefits of taking the ratio of elliptic to triangular flow, $v_2/v_3$, are discussed, and the spacetime regions which contribute dominantly to the photon flow harmonics are identified. The directed flow $v_1$ of thermal photons is predicted for RHIC and LHC energies.
\end{abstract}


\date{\today}

\maketitle

\section{Introduction}\label{sec1}

The shear viscosity of quark-gluon plasma (QGP) has received much recent attention. Experimental measurements of the collective flow of the ultradense matter created in relativistic heavy ion collisions at the Relativistic Heavy Ion Collider (RHIC) and the Large Hadron Collider (LHC) have shown that this matter exhibits almost perfect fluidity \cite{Heinz:2009xj}. The observations limit the QGP shear viscosity to entropy density ratio, $(\eta/s)_\mathrm{QGP}$, to less than $\frac{3}{4\pi}$ \cite{Heinz:2013th}, i.e. to less than 3 times the lowest value allowed by strongly-coupled conformal field theory \cite{Kovtun:2004de}. $(\eta/s)_\mathrm{QGP}$ is extracted from measured anisotropies in the flow pattern of the hadrons emitted from the collision fireball in its final freeze-out stage \cite{Heinz:2013th}. These flow anisotropies are driven by anisotropic pressure gradients in the fireball, caused by collision geometry and event-by-event fluctuations in its initial density 
distribution. Shear viscosity degrades the fluid's ability 
to convert such pressure anisotropies into flow anisotropies. It is this viscous suppression of anisotropic flow that is being exploited when extracting $(\eta/s)_\mathrm{QGP}$ from measured final state flow patterns.

All known liquids have specific shear viscosities $\eta/s$ that depend on temperature, featuring a pronounced minimum near the liquid-gas phase transition \cite{Csernai:2006zz}. The temperature dependence of the QGP shear viscosity \cite{Niemi:2012ry,Gale:2012rq} is of paramount interest since a possible rise of $(\eta/s)_\mathrm{QGP}(T)$ at temperatures above the critical value $T_c$ of the quark-hadron phase transition may indicate a gradual change of character of the QGP, from a strongly-coupled liquid near $T_c$ to a more weakly coupled gaseous plasma at much higher temperatures. The bulk of the particles created in heavy-ion collisions are strongly interacting hadrons that escape only at the very end of the fireball's evolution and thus contain only indirect information about the hottest QGP stages near the beginning of its life. Photons, on the other hand, interact only electromagnetically and can escape from the fireball during all collision stages, especially from its hot core. It was recently shown \cite{vanHees:2011vb,Shen:2013vja} that, for collision energies available at RHIC and LHC, thermal photon yields and their (azimuthally averaged) spectral slopes provide experimental information that is heavily weighted in a temperature region of ${\pm\,}50$\,MeV around the quark-hadron phase transition.
Anisotropies in the photon spectra \cite{Chatterjee:2005de}, in particular the dependence of their magnitudes $v_n$ on the harmonic order $n$, are especially sensitive to shear viscous effects. As we will show here, the largest photon $v_n$ signal also comes from the transition region, thereby reflecting the shear viscous effects in this temperature range. A measurement of anisotropic photon flow thus provides a window into fireball stages that precede those accessible through hadronic observables and complement measurements of thermal photon yields. Higher-order thermal photon flow anisotropies thus offer valuable additional constraints on the QGP specific shear viscosity.

For a fixed value of $\eta/s$, the dynamical effects from shear viscosity are proportional to the fluid's expansion rate \cite{Heinz:2009xj}. In heavy-ion collisions the expansion rate is highest at early times. Viscous effects on photon flow should therefore be more important than those for hadrons emitted at the end of the collision. This led Dusling \cite{Dusling:2009bc} to propose using photons as a QGP viscometer. We expand on this idea by studying the entire spectrum of anisotropic flows generated in the event. The anisotropic flow coefficient can be written as a complex vector, 
\begin{equation}
V_n \equiv v_n e^{i n \Psi_n} = \frac{\int p_\perp dp_\perp d\phi_p dN/(dy p_\perp dp_\perp d\phi_p) e^{i n \phi_p}}{\int p_\perp dp_\perp d\phi_p dN/(dy p_\perp dp_\perp d\phi_p)},
\label{eq1b}
\end{equation}
where $\phi_p$ is the angle of the emitted particle momentum $\bm{p}$ around the beam direction, $v_n$ is the magnitude and $\Psi_n$ is the $n$-th order flow event plane angle. We focus not just on the ``elliptic flow'' $v_2$ studied in previous work \cite{Chatterjee:2005de,Dusling:2009bc,Chatterjee:2008tp,vanHees:2011vb,Dion:2011pp,Basar:2012bp,Chatterjee:2013naa}, but also on the higher harmonic flow coefficients $v_{3,4,5}$, and on the ``directed flow'' $v_1$.

Investigating higher order flow harmonics is motivated by at least two observations: First, it is known \cite{Alver:2010dn,Schenke:2011bn} that for hadrons, shear viscosity suppresses the higher $p_T$-integrated $v_n$ coefficients  more strongly than $v_2$. Second, measurements by the PHENIX collaboration of direct photons in 200\,$A$\,GeV Au+Au collisions established a strong excess over known pQCD sources that has been attributed to thermal radiation \cite{Adare:2008ab}. The measured azimuthal anisotropy of this radiation \cite{Adare:2011zr} implies an unexpectedly large photon elliptic flow, comparable to that of pions. Recent direct photon measurements by the ALICE collaboration in 2.76\,$A$\,TeV Pb+Pb collisions at the 
LHC \cite{Wilde:2012wc,Lohner:2012ct} confirmed 
these findings which challenge our current theoretical understanding of microscopic rates and/or bulk dynamics \cite{Chatterjee:2005de,Dion:2011pp,vanHees:2011vb}. They prompted a novel idea to generate large photon elliptic flow through a non-perturbative pre-equilibrium mechanism involving the huge initial magnetic fields generated by the colliding nuclei \cite{Basar:2012bp}. Triangular photon flow $v_3$, which is purely driven by initial density fluctuations and whose direction $\Psi_3$ is therefore randomly oriented relative to the impact parameter and magnetic field \cite{Qin:2010pf,Qiu:2011iv}, should allow to disentangle the thermal photon signal from these pre-equilibrium photons. 

\section{Viscous corrections to thermal photon emission rates}\label{sec2}

In an anisotropically expanding fireball, viscosity leads to anisotropic deviations of the phase-space distribu\-tion from local equilibrium that affect thermal photon emission in two distinct ways: They alter the development of hydrodynamic flow, by increasing its radial component while suppressing its anisotropies \cite{Heinz:2009xj,Heinz:2013th}, and they modify the electromagnetic emission rate which becomes locally anisotropic. Deviations from isotropy of the local rest frame momentum distributions and anisotropic medium-induced self-energies of the exchanged quanta both contribute to this anisotropy. We here present a  study where all of these effects are included consistently, within approximations detailed below. Our approach involves the generalization of the rules of finite temperature quantum field theory to systems that are slightly out of thermal equilibrium, including all terms that are linear in the viscous correction $\pi^{\mu\nu}(x)$ to the energy momentum tensor $T^{\mu\nu}(x)$ at 
emission point $x$.

Off-equilibrium corrections to photon emission rates were studied before \cite{Schenke:2006yp,Dusling:2009bc,Dion:2011pp}, but only including the viscous corrections to the distribution functions of the particles involved in the radiation process. In the QGP, where collisions are caused by the exchange of (originally massless) gluons, dynamically generated (so-called Hard Thermal Loop (HTL)) self energies must be taken into account to regulate an infrared divergence caused by very soft collisions. In a plasma with locally anisotropic momentum distributions, these HTL self energies are anisotropic, too. In previous work \cite{Baier:1997xc,Schenke:2006fz,Schenke:2006yp}, the HTL-resummed quark self-energy was evaluated for spheroidally deformed momentum distributions. 
In Ref.~\cite{Shen:2014nfa}, whose results we now briefly summarize, we generalized the HTL resummation scheme to include anisotropic distribution functions of the more general form \cite{Dusling:2009df}
\begin{equation}
  f(p) = f_0(p) + 
  f_0(p) (1{\pm}f_0(p)) \frac{\pi^{\mu\nu} \hat{p}_\mu \hat{p}_\nu }{2(e{+}P)}\,
  \chi(p/T),
\label{eq2}
\end{equation}
where $f_0(p)$ is the Bose/Fermi equilibrium distribution function, $e$ and $P$ are the energy density and thermal pressure, $\chi(p/T) = (\vert p \vert /T)^\alpha$ with $1{\,\le\,}\alpha{\,\le\,}2$, $\hat{p}_\mu = p_\mu/\vert p \cdot u \vert$ is the momentum unit vector, and $\pi^{\mu\nu}$ is the shear stress tensor of the system. Note that $\alpha{\,=\,}2$ is used for all calculations in this paper, i.e. the viscous corrections grow approximately quadratically with the photon momentum. The effect of $\alpha$ on the photon rate is discussed in Ref.\,\cite{Shen:2014nfa}. Adding viscous corrections to the collinear emission kernel 
developed by AMY \cite{Arnold:2001ms}, necessary for a complete leading-order calculation of the viscous QGP photon emission rates, requires major theoretical development to deal with plasma instabilities, which we leave for future work.

The rate for emitting a photon in a $2{\,\to\,}2$ process $1 + 2 \rightarrow 3 + \gamma$ can be written as
\begin{eqnarray}
  E_q\frac{dR}{d^3q} &=& \frac{1}{2}\int_{\bm{p}_1, \bm{p}_2, \bm{p}_3} \!\!\!\!\!
  \vert \mathcal{M} \vert^2\, 2\pi \delta^{(4)}(p_1 + p_2 - p_3 - p_4) 
  \notag \\
&&\times f(p_1) f(p_2) (1 \pm f(p_3)),
\label{eq3}
\end{eqnarray}
where $\int_{\bm{p}}{\,\equiv\,}\frac{1}{(2\pi)^3}\int \frac{d^3p}{2 E_p}$. In the QGP phase, the $2{\,\to\,}2$ photon production channels involve quark-gluon Compton scattering and quark-antiquark annihilation. The logarithmic infrared divergence from soft collisions is regulated by using a HTL-resummed internal quark propagator \cite{Kapusta:1991qp,Baier:1991em}. The hadron gas (HG) phase is modeled as an interacting meson gas within the SU(3)${\,\times\,}$SU(3) massive Yang-Mills approach used in  previous studies (see, e.g., Refs.~\cite{Song:1993ae,Turbide:2003si,Dion:2011pp}). At tree level, this formalism contributes 8 photon-producing reaction channels involving $\pi, K, \rho, \omega$, and $a_1$ mesons \cite{Turbide:2006zz}. 

Viscous corrections are included to linear order in $\pi^{\mu\nu}$ \cite{Shen:2014nfa}, by writing the thermal photon emission rates as 
\begin{eqnarray}
  q \frac{dR}{d^3 q}(q, T) = \Gamma_0(q, T) + 
  \frac{\pi^{\mu\nu}}{2(e{+}P)} \Gamma_{\mu\nu}(q, T),
\label{eq4}
\end{eqnarray}
where $\Gamma_0$ and $\Gamma_{\mu\nu}$ represent the equilibrium contribution and the first order viscous correction to the emission rate, respectively. Since $\pi^{\mu\nu}$ is traceless and transverse to the flow velocity $u^\mu$, this can be cast into
\begin{eqnarray}
  q \frac{dR}{d^3 q} = \Gamma_0 + 
  \frac{\pi^{\mu\nu}q_\mu q_\nu}{2(e{+}P)}a_{\alpha \beta} \Gamma^{\alpha \beta}
\label{eq4a}
\end{eqnarray}
with the projection tensor 
\begin{eqnarray}
  a_{\alpha\beta} = \frac{3 q_\alpha q_\beta}{2(u{\cdot}q)^4} 
                           + \frac{u_\alpha u_\beta}{(u{\cdot}q)^2} 
                           + \frac{g_{\alpha\beta}}{2(u{\cdot}q)^2} 
                           - \frac{3(q_\alpha u_\beta{+}q_\beta u_\alpha)}{2(u{\cdot}q)^3}.\quad
\label{eq5}
\end{eqnarray}
The scalars $\pi^{\mu\nu}q_\mu q_\nu$ and $a_{\alpha \beta} \Gamma^{\alpha \beta}$ are most easily evaluated in the global and local fluid rest frames, respectively.

$\Gamma_0(q, T)$ and $\Gamma^{\alpha \beta}(q, T)$ are calculated using the decomposition (\ref{eq2}) in Eq.~(\ref{eq3}) and collecting terms independent of and linear in $\pi^{\alpha\beta}$, respectively. In the hadronic phase all internal propagators are massive, and we follow custom \cite{Song:1993ae,Turbide:2003si} by ignoring medium effects on the meson masses and coupling constants. $\pi^{\alpha\beta}$ corrections thus arise only from the explicit $f(p)$ factors for the incoming and outgoing particles in Eq.~(\ref{eq3}). In the QGP phase, the required use of HTL-resummed internal propagators to account for dynamical quark mass generation introduces a sensitivity of the matrix elements themselves to the deviation of $f(p)$ from local equilibrium: $|\mathcal{M}|^2{\,=\,}|\mathcal{M}_0|^2 + |\mathcal{M}_1|^{2,\mu\nu} \frac{\pi_{\mu\nu}}{2(e{+}P)}$. So the emission rate receives corrections ${\sim\,}\pi^{\mu\nu}$ from both the matrix elements and the explicit $f(p)$ factors. For a medium described by 
anisotropic distribution functions of the type (\ref{eq2}), the in-medium propagators continue to satisfy the KMS relation in the high temperature limit \cite{Shen:2014nfa}. We can therefore simply follow \cite{Mrowczynski:2000ed,Schenke:2006fz} to calculate the retarded quark self-energy with the modified distribution functions (\ref{eq2}).

\section{Photon flow anisotropies from event-by-event hydrodynamics}\label{sec3}

Our event-by-event simulations employ the integrated {\tt iEBE-VISHNU} framework \cite{Shen:2014vra}. The dynamical evolution of the radiating fireball is modeled with the boost-invariant hydrodynamic code {\tt VISH2+1} \cite{Song:2007fn}, using parameters extracted from previous phenomenologically successful studies of hadron production in 200\,$A$\,GeV Au+Au collisions at RHIC \cite{Shen:2010uy,Song:2010mg} and in 2.76\,$A$\,TeV Pb+Pb collisions at the LHC \cite{Shen:2011eg,Qiu:2011hf}. We explore both Monte-Carlo Glauber (MCGlb) and Monte-Carlo KLN (MCKLN) initial conditions which we propagate with $\eta/s{\,=\,}0.08$ and $\eta/s{\,=\,}0.2$, respectively \cite{Shen:2010uy,Song:2010mg,Shen:2011eg,Qiu:2011hf}, using the lattice-based equation of state (EoS) s95p-PCE-v0 \cite{Huovinen:2009yb}. This EoS implements chemical freeze-out at $T_\mathrm{chem}{\,=\,}165$\,MeV by endowing the hadrons in the hadronic phase with temperature-dependent non-equilibrium chemical potentials that keep the final stable particle ratios fixed. These chemical potentials are included in the photon emission rates from the hadronic 
phase. Hydrodynamic evolution starts at  $\tau_0{\,=\,}0.6$\,fm/$c$ and ends on an isothermal hadronic freeze-out surface of temperature $T_\mathrm{dec}{\,=\,}120$\,MeV. We compute direct photons as the sum of thermal and prompt photons. A discussion of the other possible sources of photons appears in the next section.

\begin{figure*}[ht]
  \centering
  \begin{minipage}{0.77\linewidth}
  \includegraphics[width=1.0\linewidth]{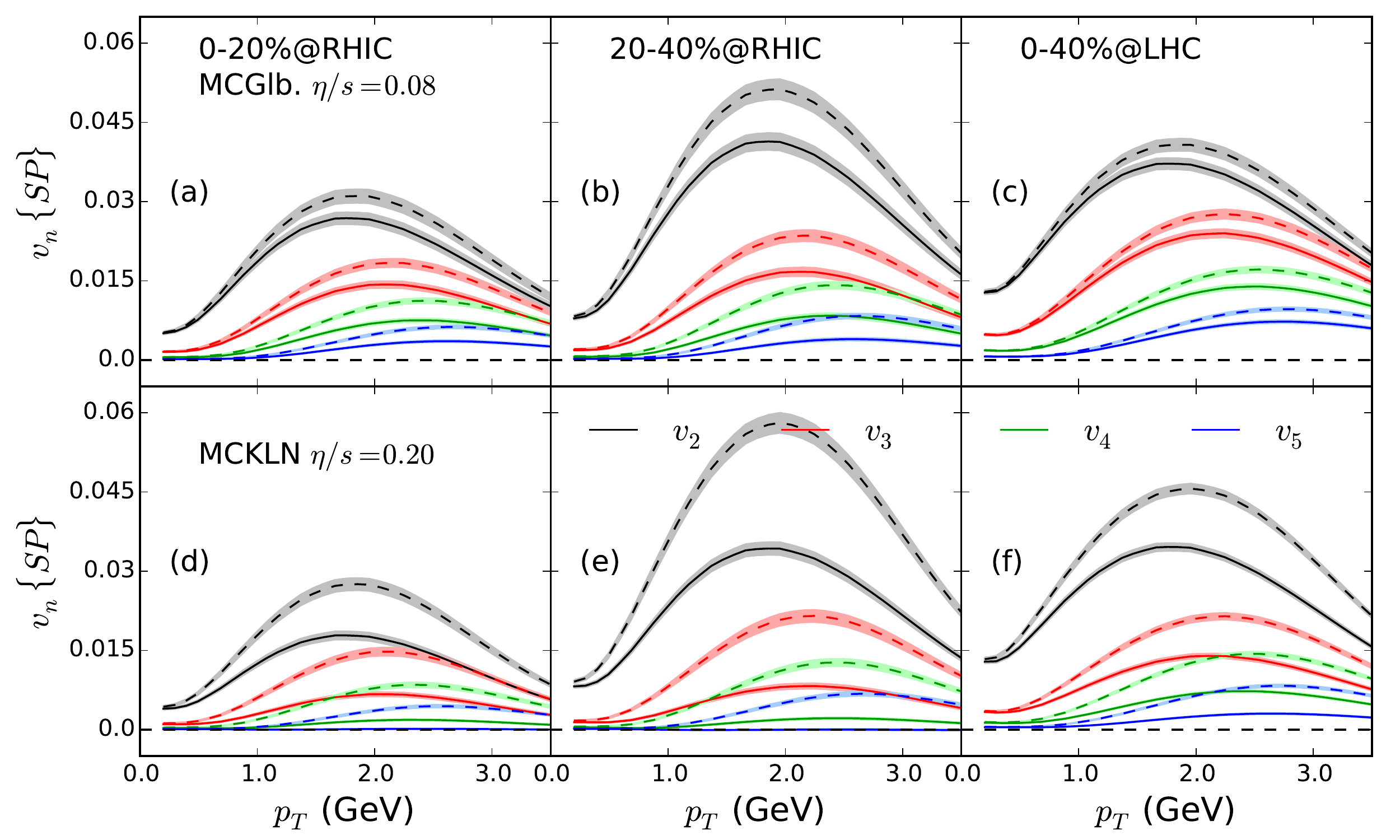}
  \end{minipage}
  \begin{minipage}{0.22\linewidth}
  \caption{(Color online) 
  Direct photon (prompt+ thermal(QGP\,{+}\,HG)) ani\-sotropic flow coefficients $v_2{-}v_5$ for 
  200\,$A$\,GeV Au+Au collisions at $0{-}20\%$ and $20{-}40\%$ centrality (left four 
  panels) and for 2.76\,$A$\,TeV Pb+Pb collisions at $0{-}40\%$ centrality (right two 
  panels). The upper (lower) row of panels shows results using MCGlb (MCKLN) 
  initial conditions with $\eta/s{\,=\,}0.08$ (0.2). Solid (dashed) lines depict results that 
  include (neglect) viscous corrections to the photon emission rates. The shaded bands 
  indicate statistical uncertainties.\\}
  \label{fig1}
  \end{minipage}
\end{figure*}

We match the thermal photon emission rate from the QGP phase to the HG phase in the temperature region, $T_\mathrm{sw} = 150-170$ MeV. The range of $T_\mathrm{sw}$ values was chosen to correspond roughly to the cross-over region of QCD at zero baryon density. There is a significant uncertainty associated with the choice of $T_\mathrm{sw}$, as shown in Ref. \cite{Shen:2014lpa}. A better understanding of photon emission in the cross-over region will be the only way of reducing this uncertainty.

The thermal photon spectrum is calculated event-by-event by folding the thermal photon emission rates with the temperature profile from the evolving hydrodynamic medium: 
\begin{equation}
   E \frac{dN^\gamma}{d^3 p} = 
   \int \tau d\tau dx dy d\eta \left(q \frac{dR}{d^3 q} (q, T)\right) 
   \bigg\vert_{q = p \cdot u(x); T(x)} \!\!\!\!\!.
\label{eq6}
\end{equation}
To this thermal spectrum for a single event we then add the prompt photon spectrum which is obtained using the smooth ensemble-averaged density profile for an average event in the same centrality class, as described in \cite{Shen:2013vja} (see discussion around Fig.~1 in that paper). The resulting total photon  spectrum -- whose ensemble-average corresponds to the sum of the red dashed and green dot-dashed lines in Fig.~1 of Ref. \cite{Shen:2013vja} where it is compared with the experimental data -- is used as weight to compute the scalar product $p_T$-differential anisotropic flow coefficients of photons $v^\gamma_n(p_T)\{\mathrm{SP}\}$ relative to the $n^\mathrm{th}$-order charged hadron flow angle $\Psi^{ch}_n$: 
\begin{eqnarray}
&& \!\!\!\!\!\!\!\!\!\!\!\!\!\! v^{\gamma}_n \{\mathrm{SP}\} (p_T) = \notag \\
&& \frac{\left \langle \frac{dN^\gamma}{dy p_T dp_T}(p_T) v^\gamma_n(p_T) v^{ch}_n \cos (n (\Psi_n^\gamma - \Psi_n^{ch}))\right \rangle}{\left \langle \frac{dN^\gamma}{dy p_T dp_T}(p_T)\right \rangle v_2^{ch}\{2\}}.
\label{eq:vnSp}
\end{eqnarray}
We use the photon multiplicity-weighted scalar product method to determine photon $v^\gamma_n\{\rm{SP}\}$ relative to the charged hadron flow planes as the method that most closely corresponds to the experimental procedure \cite{Adare:2011zr,Lohner:2012ct}. 

Results for $v_{2,3,4,5}\{\mathrm{SP}\}(p_T)$ for direct photons (thermal + prompt) from central and semi-peripheral Au+Au and Pb+Pb collisions at RHIC and LHC are shown in Fig.~\ref{fig1}.  For each centrality bin and initialization model we run 1000 fluctuating events. We emphasize that such an event-by-event approach is indispensable for the higher-order flow harmonics $n{\,\geq\,}3$, and does  influence the flow magnitude \cite{Chatterjee:2011dw,Dion:2011pp, Shen:2014cga}. Different harmonics are plotted in different colors. The difference between solid and dashed lines illustrates the importance of including viscous corrections in the emission rates; both line styles include viscous effects on the evolution of the hydrodynamic flow in the medium.\footnote{%
  Note that for MCKLN initial conditions with $\eta/s{\,=\,}0.20$, the off-equilibrium correction in the 
  photon spectrum from the term ${\sim\,}\pi^{\mu\nu}$ in Eq.~(\ref{eq2}) remains smaller than the
  equilibrium contribution up to $p_T{\,\sim\,}3.5$\,GeV in 0-40\% Pb+Pb at the LHC, and up to 
  $p_T{\,\sim\,}2.5$\,GeV in 20-40\% Au+Au at RHIC. We take those values as upper-limits on the range of $p_T$ where our results should be considered reliable.}%

Since the MCKLN-initialized fireballs are evolved with 2.5 times larger shear viscosity than the MCGlb fireballs, the viscous corrections to the emission rates are larger in the bottom panels of Fig.~\ref{fig1}. After inclusion of viscous effects, all photon $v_n$ are significantly smaller for MCKLN initial conditions than for MCGlb ones, in spite of the ${\sim\,}20\%$ larger initial ellipticity $\varepsilon_2$ from the MCKLN model \cite{Qiu:2011iv}. In a few cases, the event-plane $v_4$ and $v_5$ even become negative in the MCKLN case, driven by the large viscous corrections to the photon emission rates. 

Note that, before including viscous effects on the emission rates (dashed lines), the higher-order anisotropic flows generated from MCKLN initial conditions are larger than those from the MCGlb model, in spite of the larger $\eta/s$ used in the MCKLN runs. This is due to lower initial temperatures in hydrodynamic simulations with larger shear viscosity, in order to compensate for larger entropy production. This reduces the space-time volume for photon emission from the QGP phase and increases the ratio of photons from the hadronic phase to those from the QGP phase. Since hadronic photons carry about 10 times larger flow anisotropies, the $v_n$ of the final total photons increase. The difference between dashed and solid curves reflects the size of shear viscous suppression to the direct photon anisotropic flow from its corrections to the photon production rates. Fig.~\ref{fig1} shows that this suppression is largest between $1 \le p_T \le 3$ GeV. Although according to Eq.~(\ref{eq4a}) the viscous correction 
to thermal photon radiation increases quadratically with photon momentum, the prompt photon signal becomes dominant for $p_T > 3$ GeV, which effectively reduce the relative importance of the viscous suppression in the final direct photon anisotropic flow. This ensures our description to be well in control at high $p_T$ regions. 

%
\begin{figure*}
\centering
\includegraphics[width=0.95\linewidth]{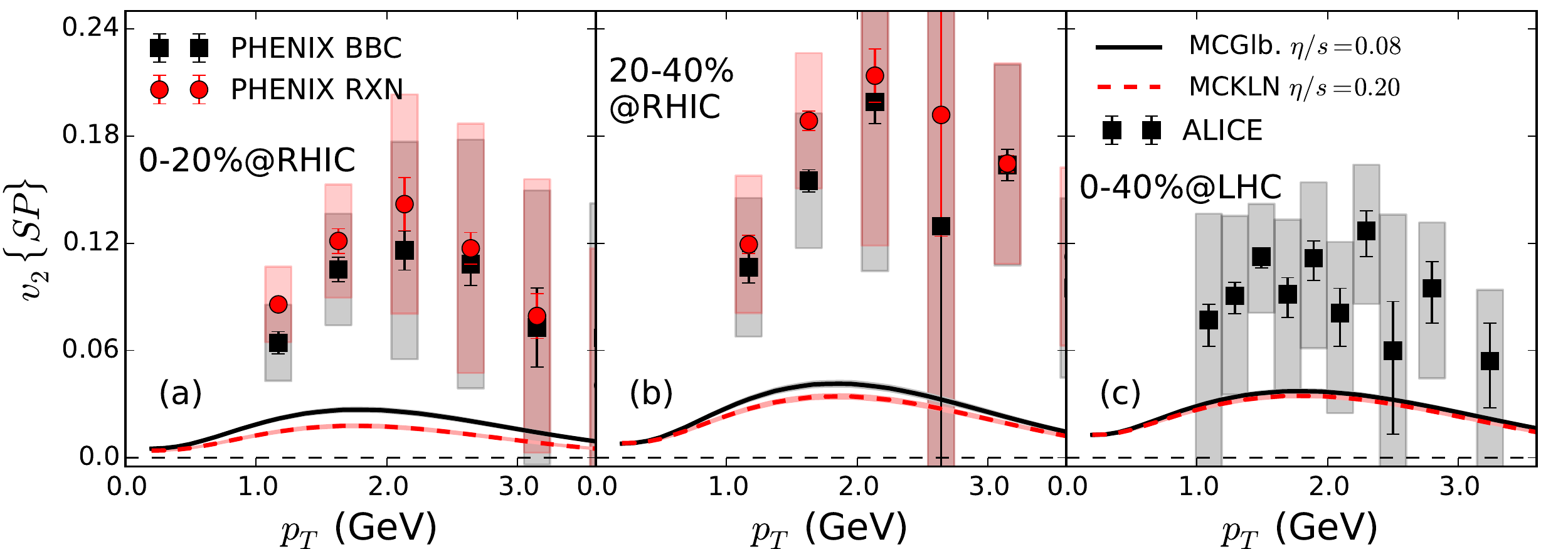}
  \caption{(Color online) Comparison of direct photon (prompt + thermal (QGP+HG)) elliptic flow 
                from event-by-event viscous hydrodynamics with recent experimental data from (a) 
                0-20\% and (b) 20-40\% central 200\,$A$\,GeV Au+Au collisions at RHIC 
                \cite{Adare:2011zr} and (c) from 0-40\% central 2.76\,$A$\,TeV Pb+Pb collisions
                at the LHC \cite{Lohner:2012ct}. Solid black (dashed red) lines correspond to MCGlb
                (MCKLN) initial conditions evolved with a shear viscosity $\eta/s{\,=\,}0.08$ (0.2), 
                respectively.  
  }
  \label{fig2}
\end{figure*}
%
%
\begin{figure*}
\centering
\includegraphics[width=0.95\linewidth]{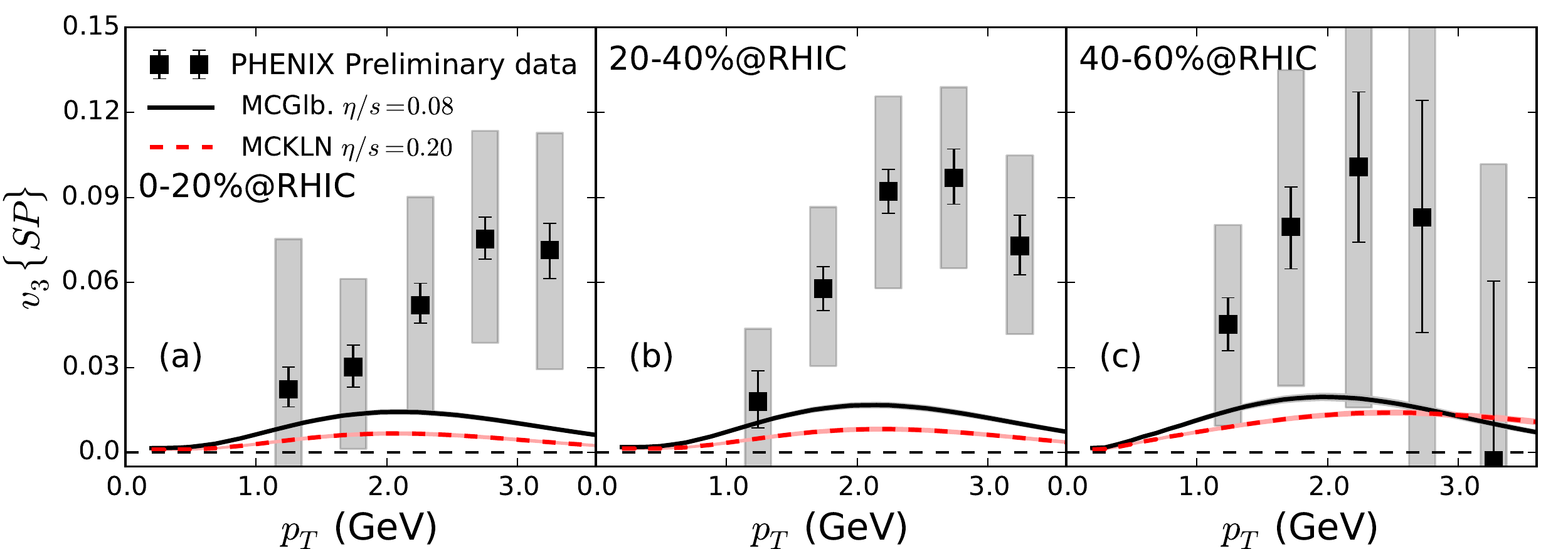}
  \caption{(Color online) Similar to Fig.~\ref{fig2} but for triangular flow.
                Direct photon (prompt + thermal (QGP+HG)) triangular flow from event-by-event 
                viscous hydrodynamics is compared with recent preliminary data from the
                PHENIX Collaboration \cite{PHENIX_preliminary_photonv3_data} for 
                200\,$A$\,GeV Au+Au collisions at RHIC, at (a) 0-20\%, (b) 20-40\%, and (c) 40-60\% 
                centrality. 
  }
  \label{fig2b}
\end{figure*}
%

The rise and fall of all $v_n$ with increasing $p_T$ reflects the dominance of hadronic photon sources (which exhibit strong flow) at low $p_T$ and the increasing weight of QGP photons from earlier and hotter stages (where flow is weak) and of prompt photons (whose anisotropic flow is assumed to vanish) at higher $p_T$ \cite{Chatterjee:2005de}. 
The slight shift of the peak of $v_n$ towards higher $p_T$ with increasing $n$ reflects the fact that the $v_n$ of the hadronic mesons which transfer their flow to the photons (pions at low $p_T$, $\rho$ and other heavier mesons at higher $p_T$) increase $\propto p_T^n$ at low $p_T$. 

Comparing central ($0{-}20\%$) to semi-peripheral ($20{-}40\%$) RHIC collisions we see that only $v_2$ increases in the more peripheral collisions, due to the increasing geometric elliptic deformation $\varepsilon_2$ of the reaction zone. The higher-order $v_n$ show little centrality dependence. A possible explanation is a cancellation between increasing hydrodynamic flow anisotropies (dashed lines) and increasing shear viscous suppression of the photon emission rate anisotropies, probably due to the smaller fireball size in peripheral collisions.

Comparing RHIC with LHC collisions we find an increase of thermal photon $v_n$ with collision energy, mainly due to the $\sim 15\%$ longer fireball lifetime at the LHC which affects mostly the QGP phase. It allows QGP photons to develop larger flow anisotropies at LHC compared to RHIC energies. The longer fireball lifetime also helps the system to evolve closer to local thermal equilibrium. The smaller ratio $\pi^{\mu\nu}/(e{+}P)$, when averaged over the fireball history, explains the smaller difference between dashed and solid lines (reflecting the photon emission rate anisotropy) at LHC energies compared to RHIC.

The direction $\Psi_n^\gamma$ of the $n^\mathrm{th}$-order photon flow is obtained by computing the phase of $\langle e^{in\phi_p}\rangle$ (where the average is taken with the $p_T$-integrated photon spectrum) \cite{Heinz:2013th}. We found that the flow angles $\Psi_n^\gamma$ for photons from the hadronic phase are tightly correlated with the charged hadron flow angles $\Psi^{ch}_n$. However, the $p_T$-dependent viscous correction to the distribution functions in Eq.~(\ref{eq2}) leads to a decorrelation between the charged hadron flow angle $\Psi^{ch}_n$ and the $p_T$-dependent photon flow angle $\Psi_n^\gamma(p_T)$ of photons with momentum $p_T$. This decorrelation increases with $p_T$ and with the shear viscosity $\eta/s$ and is largest at early times when $\pi^{\mu\nu}/(e{+}P)$ is big; it fluctuates from event to event. It becomes weaker at LHC energies where the viscous corrections are smaller.

\section{Comparison with data}\label{sec4}

In Figs.~\ref{fig2} and \ref{fig2b} we compare the differential elliptic and triangular flow coefficients $v_{2,3}^\gamma\{\mathrm{SP}\}(p_T)$ of direct photons (here, the sum of prompt and thermal photons) from our event-by-event hydrodynamic simulations with  experimental data from the PHENIX and ALICE Collaborations. One sees that both the elliptic and triangular photon flow predicted by the theoretical model falls severely short of the measured ones. Our calculations do not include pre-equilibrium photons emitted before the start of our hydrodynamic evolution at $\tau_0{\,=\,}0.6$\,fm/$c$ which presumably carry little flow anisotropy and would thus further dilute the predicted $v_2$ and $v_3$. They also do not include the viscous corrections to the soft collinear and bremsstrahlung contributions to the AMY \cite{Arnold:2001ms} thermal emission rate in the QGP phase. Since the viscous damping effects is strongest during the initial stages of the QGP phase, their inclusion is expected to further reduce the 
photon 
elliptic and triangular flow. On the other hand, we are also ignoring hadronic emission processes that involve collisions between mesons and (anti-)baryons and baryon-induced modifications of the vector meson spectral functions \cite{Turbide:2003si}, as well as meson-meson and meson-baryon bremsstrahlung processes \cite{Liu:2007zzw,Linnyk:2013wma}. Furthermore,  our current hydrodynamic approach, with its implemented sudden transition from a thermalized liquid to non-interacting, free-streaming particles does not allow for emission of photons by (increasingly rare) collisions among the dilute hadrons after kinetic freeze-out
(such collisions are included in the PHSD approach which yields better agreement with the experimental data \cite{Linnyk:2013wma}). Since all these additional hadronic photon emission processes occur during a stage where the hydrodynamic flow anisotropies have reached (most of) their final strength, their inclusion would increase the direct photon elliptic flow. Whether doing so will succeed in reproducing the experimental data remains to be seen.

\section{Photon $\mathbf{v_2/v_3}$ as a viscometer}\label{sec5}

\begin{figure}[bt]
\includegraphics[width=1.\linewidth]{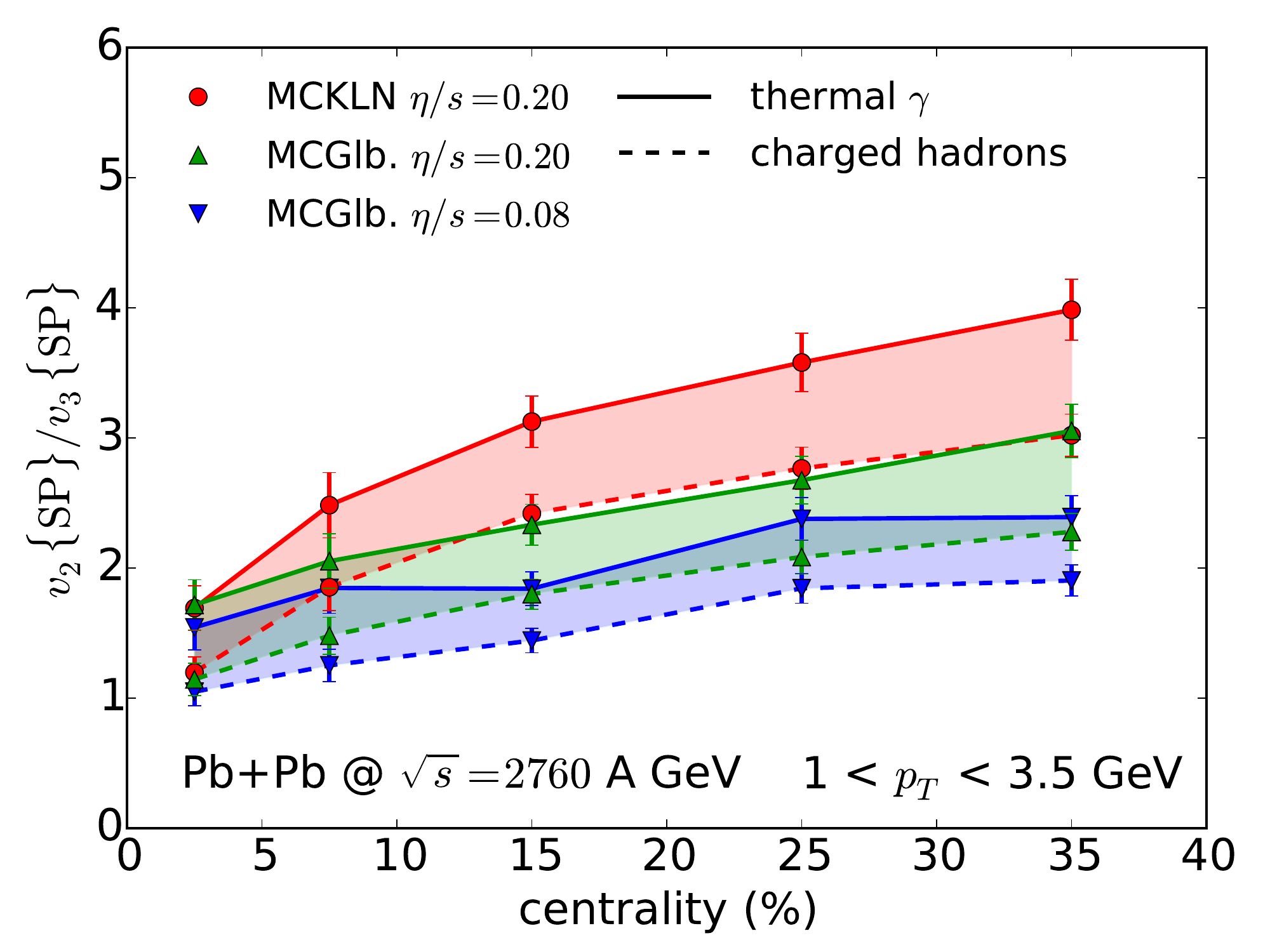}
\caption{(Color online) The ratio of the integrated elliptic flow to the integrated triangular flow, for  2.76\,$A$\,TeV Pb+Pb collisions, as a function of collision centrality. Solid and dashed lines show the ratio for thermal photons and for all charged hadrons, respectively. The ratio is shown for three pairings of initial conditions (MCGlb and MCKLN) and specific shear viscosities ($\eta/s$), as detailed in the legend. [Note that both MCGlb initial conditions with $\eta/s{\,=\,}0.08$ and MCKLN initial conditions with $\eta/s{\,=\,}0.2$ correctly reproduce the measured hadron spectra and elliptic flows while MCGlb initial conditions with $\eta/s{\,=\,}0.2$ do not \cite{Song:2010mg,Shen:2011eg}.] The shaded regions between corresponding solid and dashed lines emphasize the increase of this ratio for thermal photons.
}
\label{fig3}
\end{figure}

As was done for hadrons \cite{Qiu:2011hf}, one can form the ratio of the integrated elliptic to triangular flow coefficients, $v_2\{\mathrm{SP}\}/v_3\{\mathrm{SP}\}$, for photons and study its centrality dependence. This is shown and compared with the same ratio for all charged hadrons in Figure~\ref{fig3}. A similar analysis of the $p_T$ differential ratio $v_2\{\mathrm{SP}\}(p_T)/v_3\{\mathrm{SP}\}(p_T)$ can be found in Ref. \cite{Shen:2014cga}. 

%
%

The most interesting property of this ratio is its insensitivity to photon sources that have a vanishing $v_n$. This is not true for every definition of $v_n$; it is only the case for $v_n$ measurements where the photon $v_n$ is weighted, on an event-by-event basis, by the photon multiplicity from the same $p_T$ bin. As can be seen in Eq.~(\ref{eq:vnSp}), the consequence of this multiplicity weighting is that the multiplicity associated with zero $v_n$ photon sources only appears as a normalization, which cancels in the ratio of $v_n\{\mathrm{SP}\}(p_T)$'s. Thus this ratio has a much reduced sensitivity to photon sources like prompt and pre-equilibrium photons that are understood to carry a small $v_n$. We emphasize that this property makes the $v_2\{\mathrm{SP}\}(p_T)/v_3\{\mathrm{SP}\}(p_T)$ ratio a complementary observable to the individual $v_n\{\mathrm{SP}\}(p_T)$ measurements. An experimental measurement of this ratio for direct photons will help shed light on the dynamical flow structure prior to the hadronic kinetic freeze-out.

A study of the thermometric properties of real photons \cite{Shen:2013vja}  has shown 
that the thermal photon yield and the slope of the thermal photon spectra, while clearly reflecting fireball conditions prevailing before the emission of hadrons, do not really allow an unobstructed view of the earliest part of the fireball's expansion history. Viscous effects on the photon emission rates, on the other hand, are strongest during the earliest fireball stage when the expansion rate is largest \cite{Dusling:2009bc,Dion:2011pp,Shen:2013vja}. The  $v_2/v_3$ ratio of thermal photons, corrected for viscosity,  shown in Fig. \ref{fig3} thus provides a view of the Little Bang that is thus weighted towards earlier parts of its history than the same ratio for hadrons.

\section{Photon tomography} \label{sec6}

%
\begin{figure*}
\centering
\begin{tabular}{cc}
\includegraphics[width=0.45\linewidth]{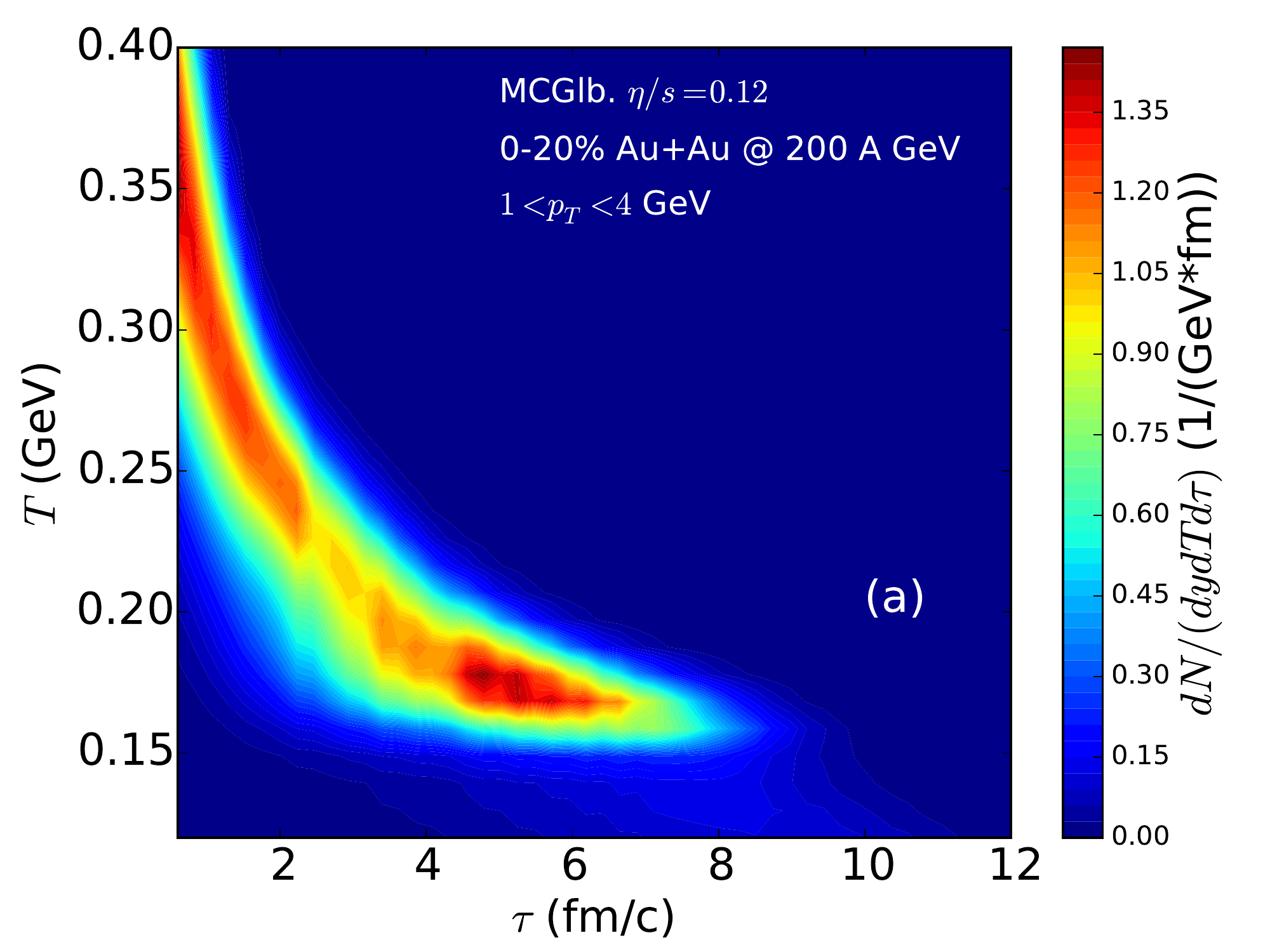} & 
\includegraphics[width=0.45\linewidth]{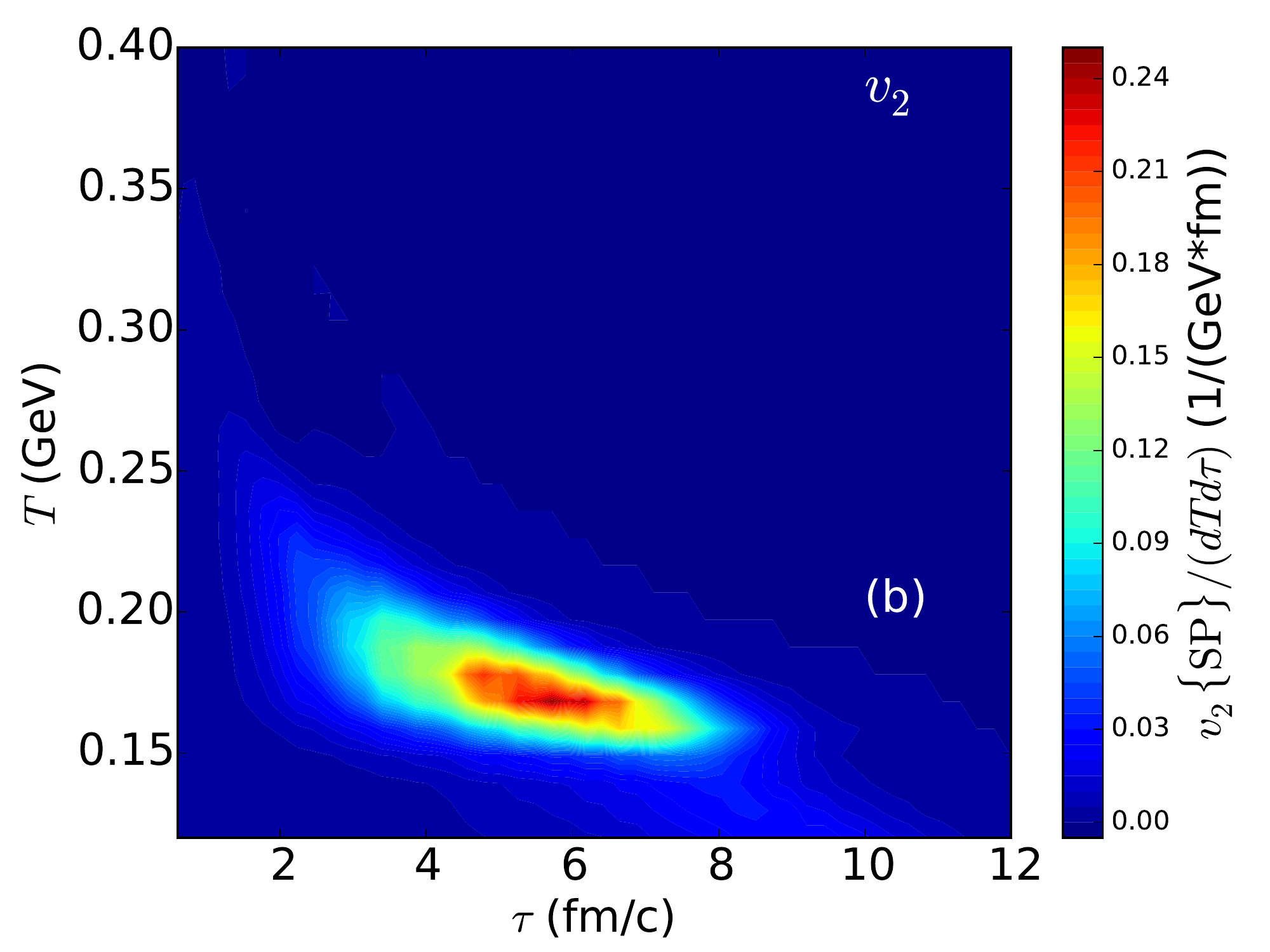} \\
\includegraphics[width=0.45\linewidth]{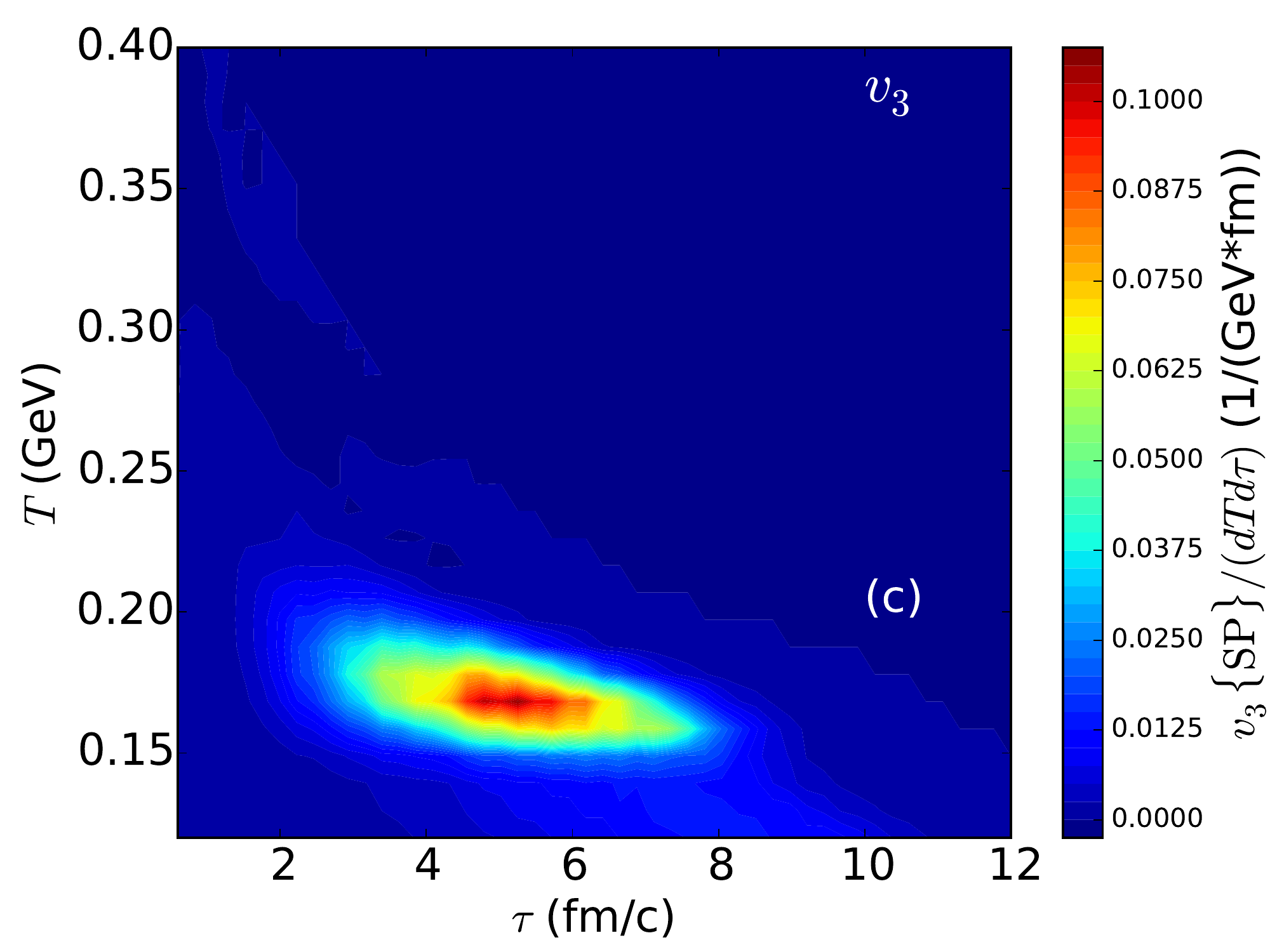} & 
\includegraphics[width=0.45\linewidth]{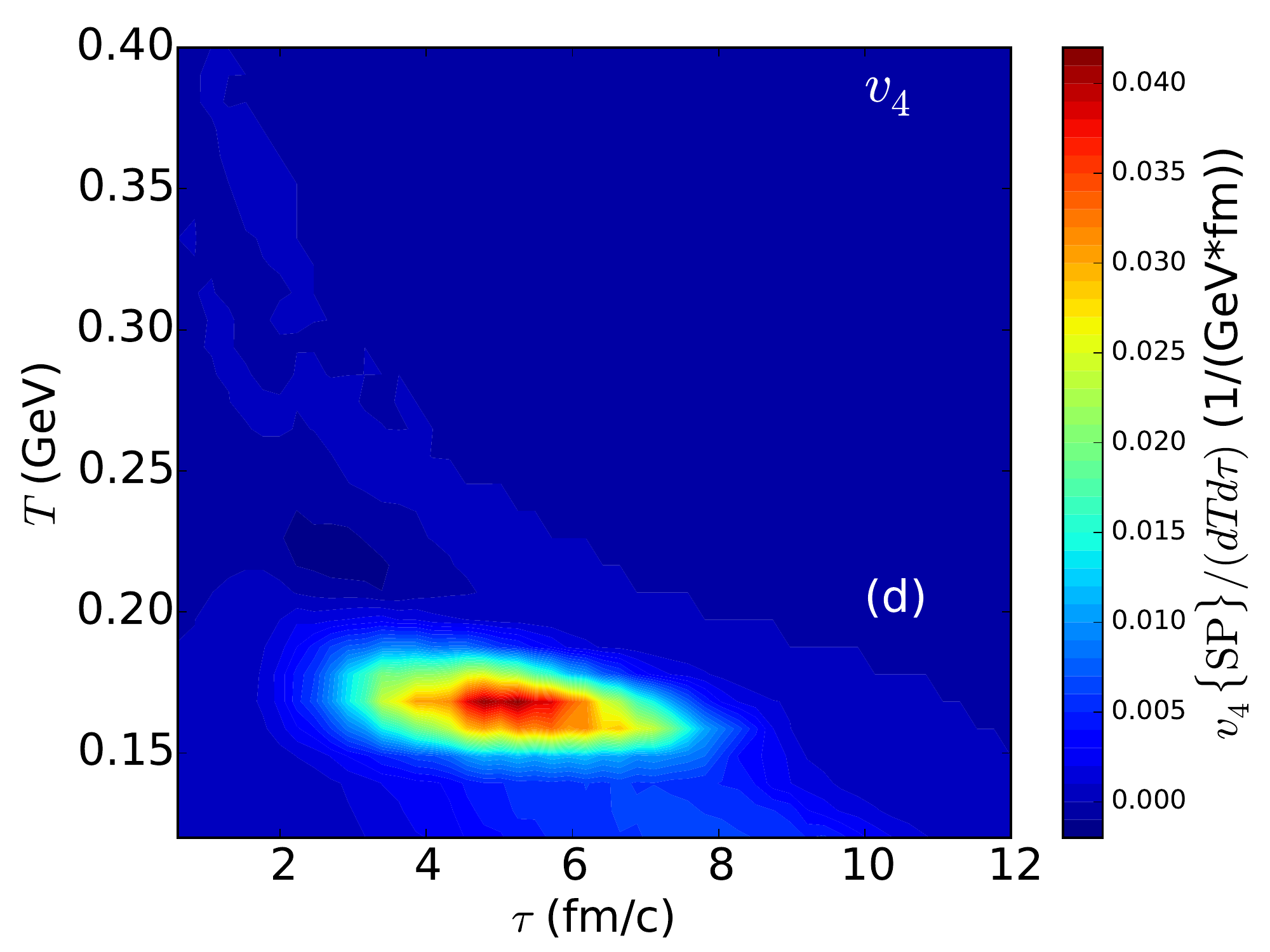}
\end{tabular}
  \caption{(Colored online) Colored contour plots for the $p_T$-integrated thermal photon emission
  yield (a) and its anisotropy coefficients, $v_2$ to $v_4$ (b-c), as functions of the local 
  temperature and emission time from event-by-event hydrodynamic simulations. All 
  observables are integrated over $1{,\leq\,}p_T{\,\leq\,}4$\,GeV/$c$.}
  \label{fig4}
\end{figure*}
%

To further study and decipher the dynamical information that is encoded in the thermal photon flow observables, we slice the hydrodynamic medium and compute the $p_T$-integrated photon emission yield as well as its anisotropic flow coefficients in Fig.\,\ref{fig4} as functions of the local temperature and emission time. We generalize the analysis in \cite{Shen:2013vja} and study 200 event-by-event hydrodynamic simulations using Monte-Carlo Glauber initial conditions with $\eta/s = 0.12$ for 0-20\% centrality in Au+Au collisions at 200 $A$ GeV.

In Figs.~\ref{fig4}b-d we show the differential contributions to the thermal photon anisotropic flow coefficients, $dv_n/(dT\,d\tau)$ for $n{\,=\,}2,3,4$, from a fluid cell $i$ at given temperature and emission time, defined as
\begin{equation}
\frac{d v_n^\gamma\{\mathrm{SP}\}(i)}{dT d\tau} = \frac{\left\langle\frac{dN_i^\gamma}{dy}\,v_n^\gamma (i)\,v_n^{ch} \cos(n (\Psi_n^\gamma(i) - \Psi_n^{ch}))\right\rangle}{\Delta T \Delta \tau \left\langle\sum_{i\in \mathrm{all\,cells}} \frac{dN_i^\gamma}{dy} \right\rangle v_2^{ch}\{2\}}.
\label{eq9}
\end{equation}
Here $\Delta T$ and $\Delta \tau$ are the small ranges of temperature and emission time that the fluid cell $i$ covers, and $\langle \cdots \rangle$ represents the average over many events. Since the $d v_n^\gamma\{\mathrm{SP}\}(i)/(dT d\tau)$ for different fluid cells share the same denominator, the definition in Eq.~(\ref{eq9}) ensures that the $d v_n^\gamma\{\mathrm{SP}\}(i)/(dT d\tau)$ integrated over the entire $T$ and $\tau$ plane will reproduce the final observed flow anisotropy of thermal photons.

Figure~\ref{fig4}a shows that thermal photon emission, although spread over the entire fireball evolution, occurs in two waves. The first wave comes from the hottest fireball region during the earliest stage of the evolution. Because at this stage hydrodynamic flow has not yet been developed, these photons only contribute to the final thermal photon yield, but not to the photon flow anisotropies. The second wave of the thermal photons comes from the temperature region near the phase transition, $T\sim$ 150-200 MeV, at times ranging approximately from $\tau = 4$ to $8$\,fm/$c$. It is due to the growth of the hydrodynamic space-time volume during the evolution, amplified by the softening of the equation of state near $T_c$. At the time of this second wave the hydrodynamic flow anisotropy has been largely developed, and the thermal photons emitted during this second wave therefore carry most of the finally observed flow anisotropies. 

%
\begin{figure*}[ht!]
\centering
\begin{tabular}{cc}
\includegraphics[width=0.45\linewidth]{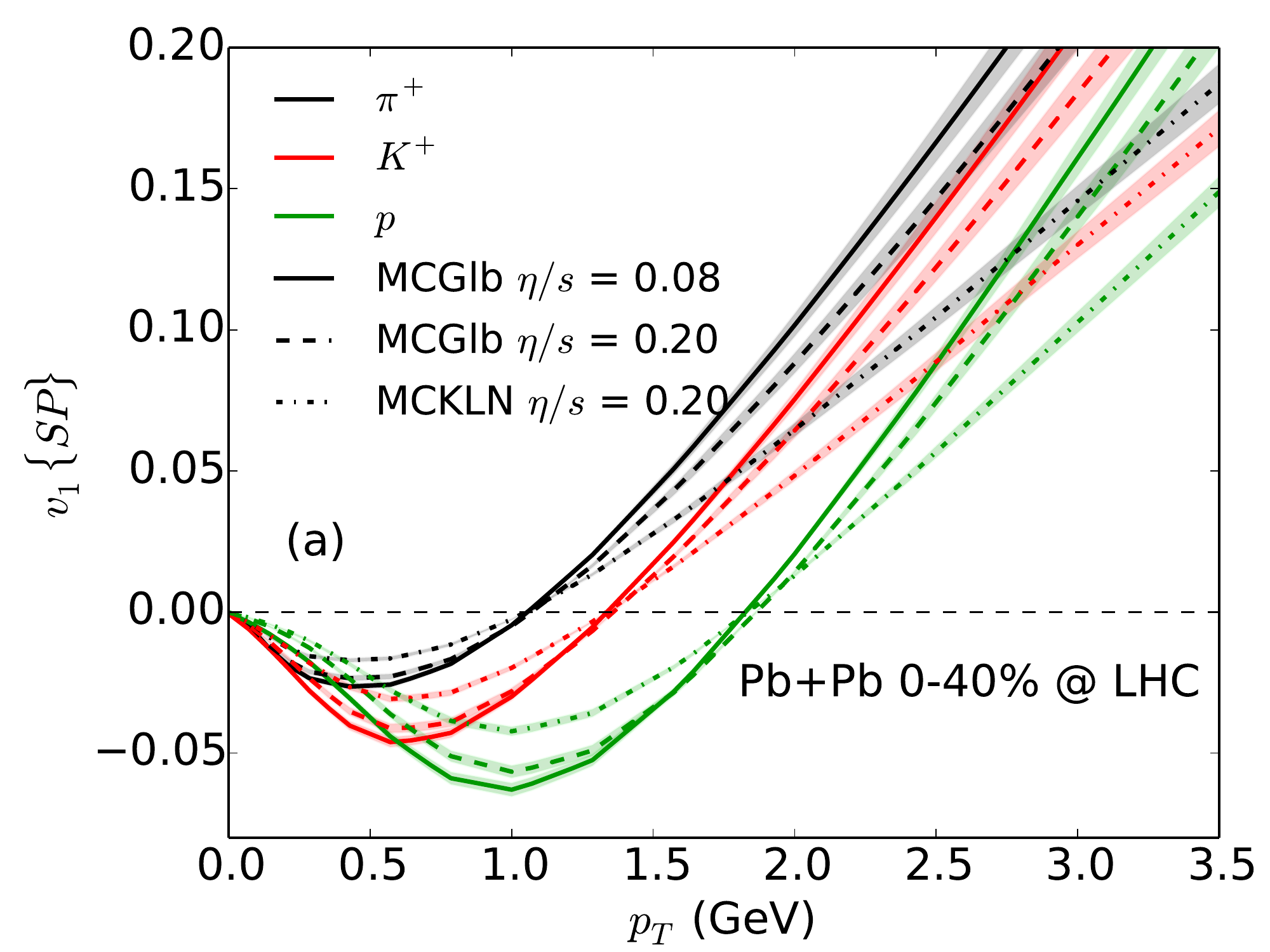} & 
\includegraphics[width=0.45\linewidth]{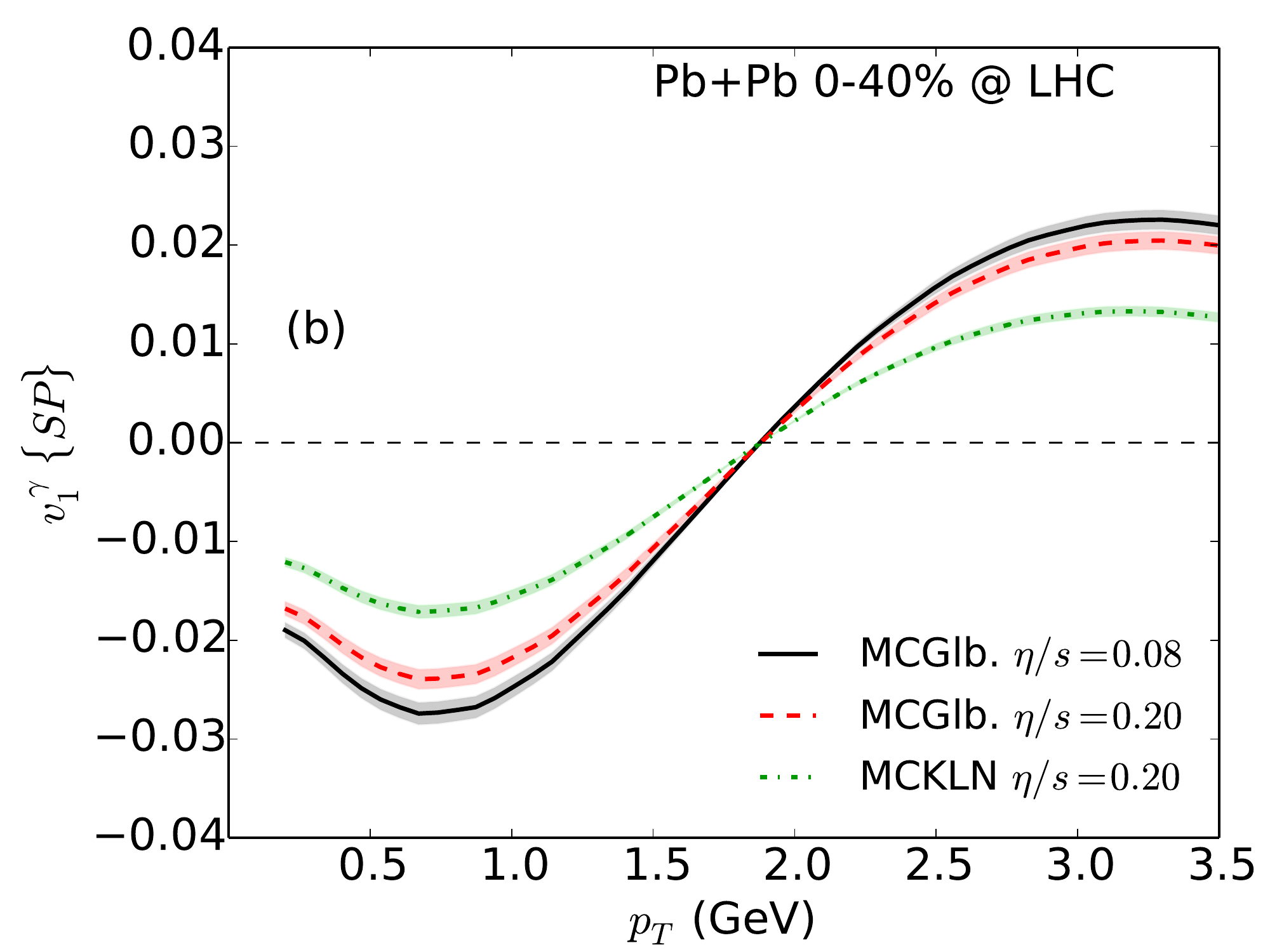} \\
\includegraphics[width=0.45\linewidth]{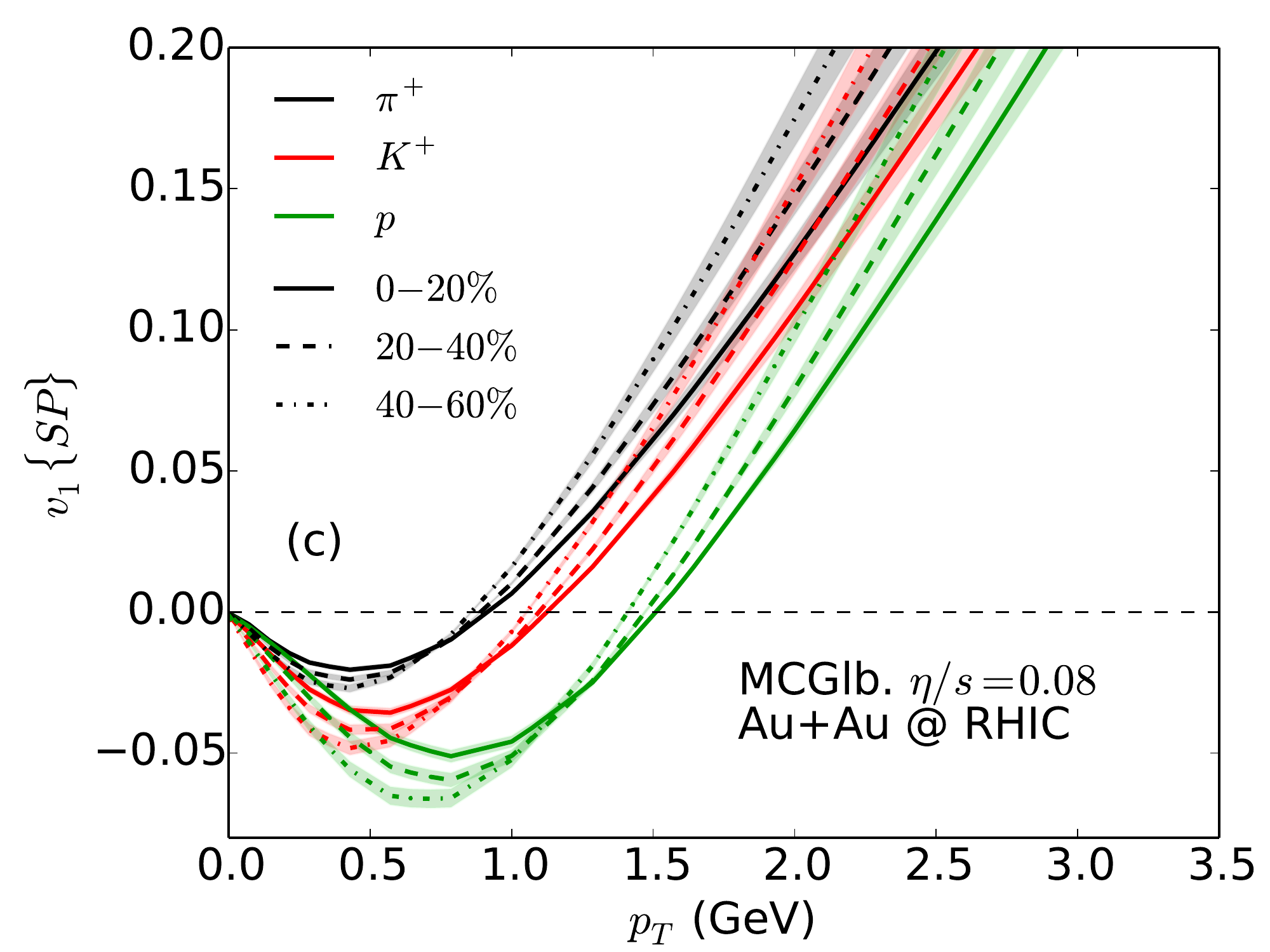} & 
\includegraphics[width=0.45\linewidth]{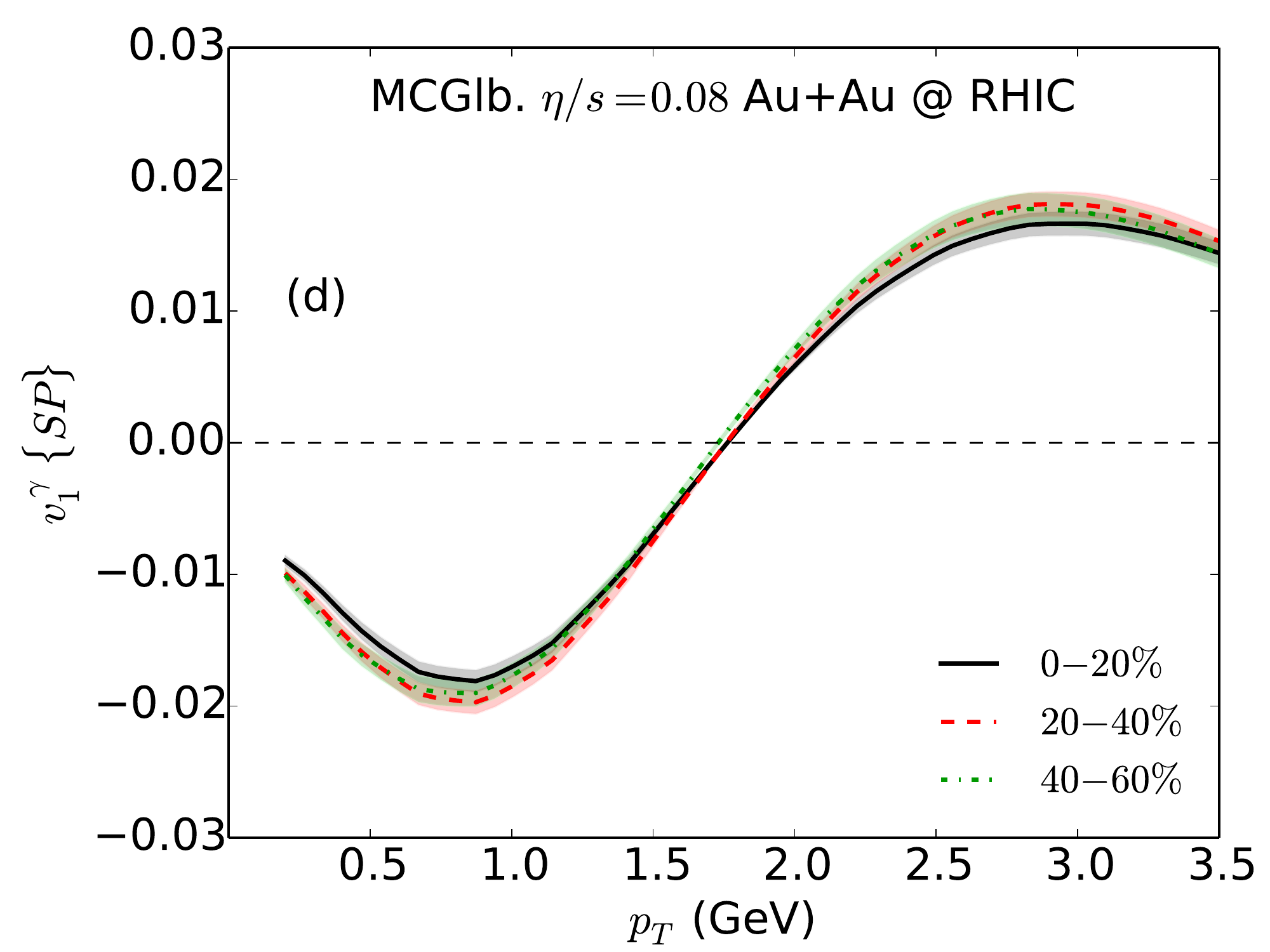} \\
\end{tabular}
\caption{(Color online)
    The scalar-product direct flow, $v_1\{\mathrm{SP}\}$, is shown for identified charged hadrons in
    panels (a, c) and for direct photons (thermal + pQCD prompt) in panels (b, d). Panels (a, b)
    show results for different initial conditions and choices of $\eta/s$, for 0-40\% central Pb+Pb 
    collisions at $\sqrt{s} = 2.76$\,$A$\,TeV. Panels (c, d) show the centrality dependence of 
    $v_1\{\mathrm{SP}\}$ in Au+Au collision at 200 A GeV at RHIC. The reference flow vector 
    $Q_1$ is integrated over $0.3{\,\leq\,}p_T{\,\leq\,}3.5$\,GeV/$c$.}
  \label{fig5}
\end{figure*}
%

The fireball regions that dominantly contribute to the $v_n$ coefficients of different harmonic order ($n{\,=\,}2,3,$ and 4) are quite similar. We emphasize that Figs.~\ref{fig4}b-d show quantitatively  that the anisotropic flow coefficients $v_n$ $(n \ge 2)$ of the penetrating direct photons provide a snapshot of the hydrodynamic flow pattern in the fireball region that is close to the phase transition. This contrasts with charged hadrons whose $v_n$ coefficients represent the hydrodynamic flow pattern at kinetic freeze-out. The direct photon $v_n$ thus provide us with valuable information about the dynamical evolution of the fireball that complements that from the anisotropic flows of hadrons. A combined flow analysis of charged or identified hadrons and direct photons can therefore help to constrain the evolution of relativistic heavy-ion collisions more tightly than would be possible with hadronic observables alone. 

\section{Direct photon $v_1$} \label{sec7}

We close our discussion by proposing the measurement of a new electromagnetic probe in relativistic heavy-ion collisions, the direct photon directed flow $v_1$. Unlike $V_n (n \ge 2)$, the direct flow component $V_1$ is constrained by global momentum conservation. In order to separate the underlying collective behavior from correlations due to momentum conservation we follow the prescription developed by Luzum and Ollitrault \cite{Luzum:2010fb,Gardim:2011qn,Retinskaya:2012ky}: For each event we define the flow vector $Q_1$ for the measurement of directed flow by
\begin{equation}
Q_1 \equiv q_1 e^{i \tilde{\Psi}_1} = \int p_T dp_T d\phi \left(p_T - \frac{\langle p_T^2 \rangle}{\langle p_T \rangle} \right) \frac{d N}{dy p_T dp_T d\phi} e^{i \phi}.
\label{eq11}
\end{equation}
Here $\langle p_T \rangle$ and $\langle p_T^2 \rangle$ are the mean values of $p_T$ and $p_T^2$, and the $p_T$-integration runs over the desired $p_T$ range.

To measure $v_1$ of photons we correlate each photon with a reference flow vector $Q_1^\mathrm{ch}$ for all charged hadrons from the same event. The scalar-product $v_1$ for direct photons is defined by
\begin{eqnarray}
&& \!\!\!\!\!\!\!\!\!\!\!\!\!\!\!\! v_1^\gamma\{\mathrm{SP}\}(p_T)   \notag \\
&=& \frac{\left\langle \frac{dN^\gamma}{dy p_T dp_\perp}(p_T) v^\gamma_1(p_T) q_1^\mathrm{ch} \cos (\Psi_1^\gamma(p_T) - \tilde{\Psi}_1^\mathrm{ch}) \right\rangle}{\left\langle \frac{dN^\gamma}{dy p_T dp_T}(p_T) \right\rangle \sqrt{\left\langle Q_1^\mathrm{ch} \cdot (Q_1^\mathrm{ch})^*\right\rangle}} \notag \\
&=& \frac{\left\langle \frac{dN^\gamma}{dy p_T dp_T}(p_T) V^\gamma_1(p_T) \cdot (Q_1^\mathrm{ch})^*\right\rangle}{\left\langle \frac{dN^\gamma}{dy p_T dp_T}(p_T) \right\rangle \sqrt{\left\langle Q_1^\mathrm{ch} \cdot (Q_1^\mathrm{ch})^*\right\rangle}}.
\end{eqnarray}
Here $\langle\dots\rangle$ again denotes the average over collision events.

%
\begin{figure}[b!]
\centering
\begin{tabular}{cc}
\includegraphics[width=1.0\linewidth]{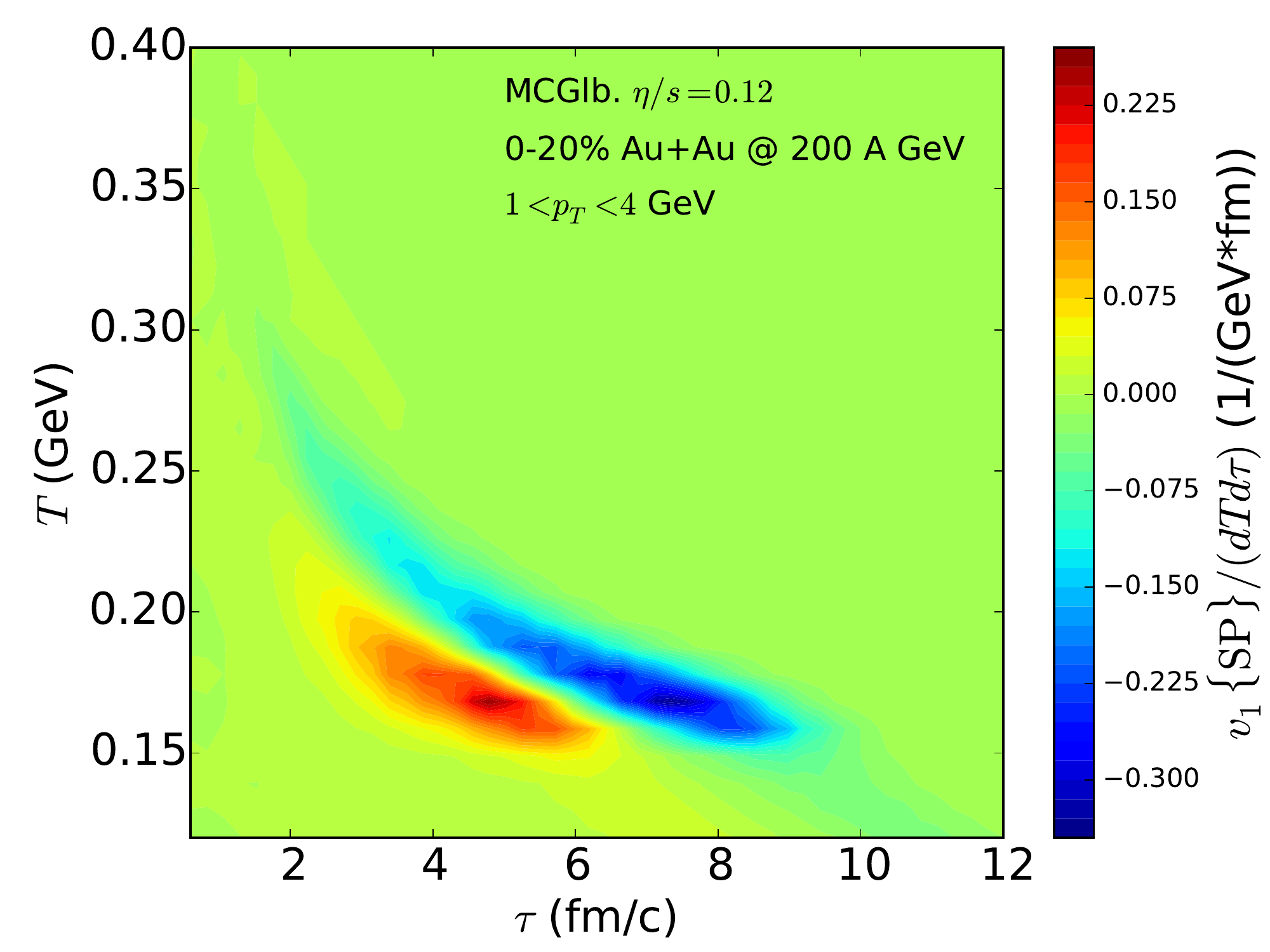}
\end{tabular}
  \caption{Colored contour plot for the $p_T$-integrated direct flow of thermal photons at different temperatures and emission times. The direct flow $v_1^\gamma\{SP\}$ is integrated over $1 \le p_T \le 4$ GeV. }
  \label{fig6}
\end{figure}
%

Figure~\ref{fig5} shows the scalar-product directed flow for identified charged hadrons (a,c) and for direct photons (b,d). The characteristic feature of negative $v_1\{\mathrm{SP}\}$ at low $p_T$ turning positive at high $p_T$ is a consequence of momentum conservation, $\int p^2_T dp_T \frac{dN}{dy p_T dp_T}\, e^{i\phi}{\,=\,}0$. The zero crossing point is largely controlled by the momentum conservation constraint and the shape of the single particle spectra. Flatter particle spectra result in larger $p_T$ values for the zero crossing of $v_1\{\mathrm{SP}\}(p_T)$. The slope of the $p_T$-differential $v_1\{\mathrm{SP}\}$ at the crossing point depends primarily on the absolute value of the $p_T$-integrated $v_1\{\mathrm{SP}\}$. In Fig.~\ref{fig5}a we clearly see a mass ordering of $v_1$ among different hadron species. As for the higher harmonics, it is caused by radial flow which flattens the $p_T$ spectra by a blueshift factor that, at low $p_T$, increases with the mass of the hadrons. Note, however, that 
the directed flow $v_1^\gamma\{\mathrm{SP}\}$ of the massless photons, shown in Fig.~\ref{fig5}b, does not follow this mass ordering. By comparing the solid and dashed curves in panels a and b we also see that a larger specific shear viscosity results in a smaller slope of $v_1\{\mathrm{SP}\}(p_T)$. This reflects the shear viscous damping of the anisotropic flow coefficient: larger $\eta/s$ reduce the $p_T$-integrated $v_1$. This observation is true for both hadrons and direct photons. A much larger difference in the slope of $v_1\{\mathrm{SP}\}(p_T)$ is observed if different initial conditions are used: MCKLN initial conditions results in a smaller directed flow coefficient than MCGlb initial conditions. 

In Figs.~\ref{fig5}c,d we further study the centrality dependence of the identified hadron and direct photon $v_1$ at the top RHIC energy. In contrast to the directed flow of charged hadrons, direct photon $v_1$ shows almost no centrality dependence. We checked that this is mostly due to the dilution from pQCD prompt photons, especially at high $p_T$.

Similar to Sec.~\ref{sec6}, we have performed a photon tomography analysis also for the thermal  photon $v_1\{\mathrm{SP}\}$. As seen in Fig.~\ref{fig6}, the differential contributions to the directed flow of photons feature more interesting structures than those to the higher order harmonic flow coefficients shown in Figs.~\ref{fig4}b-d. Similar to the higher order $v_n (n \ge 2)$, the largest  $v_1$ signal is coming from the region near the phase transition, at temperatures $T \sim 150 - 200$\,MeV. However, as a function of time, the $p_T$-integrated directed photon flow undergoes an oscillation: fireball regions that reach the transition temperature early contribute a positive directed flow while those regions that hadronize later contribute negative photon directed flow. To understand this phenomenon we note that the flow angle associated with the reference flow vector $Q_1^\mathrm{ch}$ is aligned with the differential charged hadron directed $V_1(p_T)$ at high $p_T$. These hadrons are on average emitted 
earlier than low-$p_T$ hadrons. Thermal photons emitted during the early evolution of the fireball reflect a flow pattern that is mostly parallel to that of high $p_T$ hadrons, giving a positive directed flow  signal. Photons that are emitted later carry opposite $v_1$ because their flow is mostly correlated with that of low-$p_T$ hadrons. It is  important to emphasize that the directed flow of thermal photons therefore carries  precious dynamical information that is  absent from the other photon flow harmonic coefficients.

\section{Summary}\label{sec8}

We have presented the first viscous calculation for higher order anisotropic flows for thermal photons. We find sizable triangular flow $v_3$ for thermal photons at both RHIC and LHC energies which (by symmetry) cannot be due to the initial magnetic field. Viscous effects on the anisotropic flows of thermal photons are larger than for hadrons, due to large viscous anisotropies in the photon emission rates, especially at early times. A comparison of $v_2/v_3$ for thermal photons and pions as a function of collision centrality is shown to provide a novel handle on the QGP shear viscosity that has the potential to critically complement the information extractable from hadronic anisotropic flow effects. This makes thermal photons an important additional probe of the QGP shear viscosity. Based on our event-by-event hydrodynamic simulations, we pinpointed the fireball space-time regions that contribute most of the flow anisotropy of thermal photon. In contrast to charged hadron $v_n$, 
thermal photons imprint the anisotropic flow pattern near and slightly above the cross-over phase transition region. Thus the direct photon anisotropic flows are sensitive to the phase transition region and can serve as a direct probe to constrain its theoretical modeling. We also propose a measurement of direct photon $v_1$ as a new and interesting electromagnetic flow signature, and provide predictions for this observable at both RHIC and LHC energies. 

\acknowledgments{This work was supported by the U.S. Department of Energy under Grants No. \rm{DE-SC0004286} and (within the framework of the JET Collaboration) \rm{DE-SC0004104}, and by the Natural Sciences and Engineering Research Council of Canada. The authors acknowledge useful conversations with Gabriel Denicol, Jacopo Ghiglieri, Sang\-yong Jeon, Yuri Kovchegov, Matthew Luzum, Guy Moore, Zhi Qiu, Ralf Rapp, Vladimir Skokov, and \mbox{Gojko} Vujanovic. J.-F.P. acknowledges support through grants from Hydro-Quebec and from FRQNT, and I.K. from the Canadian Institute of Nuclear Physics.

\newpage



\begin{thebibliography}{99}

\bibitem{Heinz:2009xj} 
  U.~Heinz,
  in {\sl Relativistic Heavy Ion Physics}, Landolt-Boernstein New Series, I/23, edited by 
  R. Stock (Springer Verlag, New York, 2010) Chap. 5
  [arXiv:0901.4355 [nucl-th]].

\bibitem{Heinz:2013th} 
  U.~Heinz and R.~Snellings,
  Annu. Rev. Nucl. Part. Sci. {\bf 63}, 123 (2013).

\bibitem{Kovtun:2004de} 
  P.~Kovtun, D.~T.~Son and A.~O.~Starinets,
  Phys.\ Rev.\ Lett.\  {\bf 94}, 111601 (2005).
  
\bibitem{Csernai:2006zz} 
  L.~P.~Csernai, J.~.I.~Kapusta and L.~D.~McLerran,
  Phys.\ Rev.\ Lett.\  {\bf 97}, 152303 (2006).

\bibitem{Niemi:2012ry} 
  H.~Niemi, G.~S.~Denicol, P.~Huovinen, E.~Molnar and D.~H.~Rischke,
  Phys.\ Rev.\ C {\bf 86}, 014909 (2012). 
 
 \bibitem{Gale:2012rq} 
  C.~Gale, S.~Jeon, B.~Schenke, P.~Tribedy and R.~Venugopalan,
  Phys.\ Rev.\ Lett.\  {\bf 110}, 012302 (2013).

\bibitem{vanHees:2011vb} 
  H.~van Hees, C.~Gale and R.~Rapp,
  Phys.\ Rev.\ C {\bf 84}, 054906 (2011).
  
\bibitem{Shen:2013vja} 
  C.~Shen, U.~Heinz, J.~F.~Paquet and C.~Gale,
  Phys.\ Rev.\ C {\bf 89}, 044910 (2014).

\bibitem{Chatterjee:2005de} 
  R.~Chatterjee, E.~S.~Frodermann, U.~Heinz and D.~K.~Srivastava,
  Phys.\ Rev.\ Lett.\  {\bf 96}, 202302 (2006);
  U.~Heinz, R.~Chatterjee, E.~S.~Frodermann, C.~Gale and D.~K.~Srivastava,
  Nucl.\ Phys.\ A {\bf 783}, 379 (2007).
  
\bibitem{Dusling:2009bc} 
  K.~Dusling,
  Nucl.\ Phys.\ A {\bf 839}, 70 (2010).

\bibitem{Chatterjee:2008tp} 
  R.~Chatterjee and D.~K.~Srivastava,
  Phys.\ Rev.\ C {\bf 79}, 021901 (2009).

\bibitem{Dion:2011pp} 
  M.~Dion, J.-F.~Paquet, B.~Schenke, C.~Young, S.~Jeon and C.~Gale,
  Phys.\ Rev.\ C {\bf 84}, 064901 (2011).
  
\bibitem{Basar:2012bp} 
  G.~Basar, D.~Kharzeev and V.~Skokov,
  Phys.\ Rev.\ Lett.\  {\bf 109}, 202303 (2012).

\bibitem{Chatterjee:2013naa} 
  R.~Chatterjee, H.~Holopainen, I.~Helenius, T.~Renk and K.~J.~Eskola,
  Phys.\ Rev.\ C {\bf 88}, 034901 (2013).

\bibitem{Alver:2010dn} 
  B.~H.~Alver, C.~Gombeaud, M.~Luzum and J.-Y.~Ollitrault,
  Phys.\ Rev.\ C {\bf 82}, 034913 (2010).

\bibitem{Schenke:2011bn} 
  B.~Schenke, S.~Jeon and C.~Gale,
  Phys.\ Rev.\ C {\bf 85}, 024901 (2012).
  
\bibitem{Adare:2008ab} 
  A.~Adare {\it et al.}  [PHENIX Collaboration],
  Phys.\ Rev.\ Lett.\  {\bf 104}, 132301 (2010).

\bibitem{Adare:2011zr} 
  A.~Adare {\it et al.}  [PHENIX Collaboration],
  Phys.\ Rev.\ Lett.\  {\bf 109}, 122302 (2012).
   
\bibitem{Wilde:2012wc} 
  M.~Wilde {\it et al.} (ALICE Collaboration),
  Nucl.\ Phys.\ {\bf A904-905}, 573c (2013).

\bibitem{Lohner:2012ct} 
  D.~Lohner [ALICE Collaboration],
  J.\ Phys.\ Conf.\ Ser.\  {\bf 446}, 012028 (2013).
 
\bibitem{Qin:2010pf} 
  G.-Y.~Qin, H.~Petersen, S.~A.~Bass and B.~Muller,
  Phys.\ Rev.\ C {\bf 82}, 064903 (2010).
 
\bibitem{Qiu:2011iv} 
  Z.~Qiu and U.~Heinz,
  Phys.\ Rev.\ C {\bf 84}, 024911 (2011).

\bibitem{Schenke:2006yp} 
  B.~Schenke and M.~Strickland,
  Phys.\ Rev.\ D {\bf 76}, 025023 (2007).
  
\bibitem{Baier:1997xc} 
  R.~Baier, M.~Dirks, K.~Redlich and D.~Schiff,
  Phys.\ Rev.\ D {\bf 56}, 2548 (1997).
  
\bibitem{Schenke:2006fz} 
  B.~Schenke and M.~Strickland,
  Phys.\ Rev.\ D {\bf 74}, 065004 (2006).
  
\bibitem{Shen:2014nfa} 
  C.~Shen, J.~F.~Paquet, U.~Heinz and C.~Gale,
  arXiv:1410.3404 [nucl-th].
    
\bibitem{Dusling:2009df} 
  K.~Dusling, G.~D.~Moore and D.~Teaney,
  Phys.\ Rev.\ C {\bf 81}, 034907 (2010).

\bibitem{Arnold:2001ms} 
  P.~B.~Arnold, G.~D.~Moore and L.~G.~Yaffe,
  JHEP {\bf 0112}, 009 (2001).
  
\bibitem{Turbide:2007mi} 
  S.~Turbide, C.~Gale, E.~Frodermann and U.~Heinz,
  Phys.\ Rev.\ C {\bf 77}, 024909 (2008).
  
\bibitem{Song:1993ae} 
 C.~Song,
 Phys.\ Rev.\ C {\bf 47}, 2861 (1993).

\bibitem{Turbide:2003si} 
  S.~Turbide, R.~Rapp and C.~Gale,
  Phys.\ Rev.\ C {\bf 69}, 014903 (2004).
  
\bibitem{Turbide:2006zz} 
 S.~Turbide,
 PhD thesis, McGill University, 2006.

\bibitem{Kapusta:1991qp} 
  J.~I.~Kapusta, P.~Lichard and D.~Seibert,
  Phys.\ Rev.\ D {\bf 44}, 2774 (1991)
  [Erratum: Phys. Rev.\ D {\bf 47}, 4171 (1993)].

\bibitem{Baier:1991em} 
  R.~Baier, H.~Nakkagawa, A.~Niegawa and K.~Redlich,
  Z.\ Phys.\ C {\bf 53}, 433 (1992).
  
\bibitem{Mrowczynski:2000ed} 
  S.~Mrowczynski and M.~H.~Thoma,
  Phys.\ Rev.\ D {\bf 62}, 036011 (2000).

\bibitem{Shen:2014vra} 
  C.~Shen, Z.~Qiu, H.~Song, J.~Bernhard, S.~Bass and U.~Heinz,
  arXiv:1409.8164 [nucl-th].
 
\bibitem{Song:2007fn} 
  H.~Song and U.~Heinz,
  Phys.\ Lett.\ B {\bf 658}, 279 (2008);
  Phys.\ Rev.\ C {\bf 77}, 064901 (2008);
  and  
  Phys.\ Rev.\ C {\bf 78}, 024902 (2008).
  
\bibitem{Shen:2010uy} 
  C.~Shen, U.~Heinz, P.~Huovinen and H.~Song,
  Phys.\ Rev.\ C {\bf 82}, 054904 (2010).

\bibitem{Song:2010mg} 
  H.~Song, S.~A.~Bass, U.~Heinz, T.~Hirano and C.~Shen,
  Phys.\ Rev.\ Lett.\  {\bf 106}, 192301 (2011)
  [Erratum: Phys. Rev. Lett.\  {\bf 109}, 139904 (2012)];
%
  Phys.\ Rev.\ C {\bf 83}, 054910 (2011)
  [Erratum: Phys. Rev.\ C {\bf 86}, 059903 (2012)].

\bibitem{Shen:2011eg} 
  C.~Shen, U.~Heinz, P.~Huovinen and H.~Song,
  Phys.\ Rev.\ C {\bf 84}, 044903 (2011).

\bibitem{Qiu:2011hf} 
  Z.~Qiu, C.~Shen and U.~Heinz,
  Phys.\ Lett.\ B {\bf 707}, 151 (2012).

\bibitem{Huovinen:2009yb} 
  P.~Huovinen and P.~Petreczky,
  Nucl.\ Phys.\ A {\bf 837}, 26 (2010).
  
\bibitem{Shen:2014lpa} 
  C.~Shen, J.~F.~Paquet, J.~Liu, G.~Denicol, U.~Heinz and C.~Gale,
  Nucl.\ Phys.\ A {\bf 931}, 675 (2014).
  
\bibitem{Chatterjee:2011dw} 
  R.~Chatterjee, H.~Holopainen, T.~Renk and K.~J.~Eskola,
  Phys.\ Rev.\ C {\bf 83}, 054908 (2011).

\bibitem{Shen:2014cga} 
  C.~Shen, U.~Heinz, J.~F.~Paquet and C.~Gale,
  Nucl. Phys. A {\bf 932}, 184 (2014).

\bibitem{PHENIX_preliminary_photonv3_data}
   The preliminary PHENIX data were extracted from Takao Sakaguchi's talk at the 2014 
   RIKEN BNL Research Center Workshop {\it Thermal Photons and Dileptons in 
   Heavy-Ion Collisions}.

\bibitem{Liu:2007zzw} 
  W.~Liu and R.~Rapp,
  Nucl.\ Phys.\ A {\bf 796}, 101 (2007).

\bibitem{Linnyk:2013wma} 
  O.~Linnyk, W.~Cassing and E.~Bratkovskaya,
  Phys.\ Rev.\ C {\bf 89}, 034908 (2014).

\bibitem{Heinz:2014uga} 
  U.~Heinz, J.~Liu and C.~Shen,
  Nucl. Phys. A {\bf 932}, 310 (2014).

\bibitem{Heinz:2013bua} 
  U.~Heinz, Z.~Qiu and C.~Shen,
  Phys.\  Rev.\  C 87, {\bf 034913} (2013).

\bibitem{Luzum:2010fb} 
  M.~Luzum and J.~Y.~Ollitrault,
  Phys.\ Rev.\ Lett.\  {\bf 106}, 102301 (2011).
  
\bibitem{Gardim:2011qn} 
  F.~G.~Gardim, F.~Grassi, Y.~Hama, M.~Luzum and J.~Y.~Ollitrault,
  Phys.\ Rev.\ C {\bf 83}, 064901 (2011).
  
\bibitem{Retinskaya:2012ky} 
  E.~Retinskaya, M.~Luzum and J.~Y.~Ollitrault,
  Phys.\ Rev.\ Lett.\  {\bf 108}, 252302 (2012).


\end{thebibliography}

\end{document}